\documentclass{article}
\usepackage{emulateapj,pstricks,apjfonts}

\def\maximai{{\sc maxima-i}}

\def\boom{{\sc boomerang-ldb}}

\def\letter{{\it Letter}}

\def\simlt{\lower.5ex\hbox{$\; \buildrel < \over \sim \;$}}
\def\simgt{\lower.5ex\hbox{$\; \buildrel > \over \sim \;$}}
  
\def\eg{{\rm e.g.}}
\def\ie{{\rm i.e.}}
\def\cf{{\rm cf.}}

\def\etal{{\rm et al.\ }}

\def\l#1{\left#1}
\def\r#1{\right#1}

\def\apj{{\sl Ap.J.}}

\def\mnras{{\sl MNRAS}}
\def\nat{{\sl Nature}}
\def\prl{{\sl Phys.\ Rev.\ Lett.}}
\def\prd{{\sl Phys.\ Rev.\ D.}}

\begin{document}


\submitted{The Astrophysical Journal, 561, L7-L10, 2001, November 1}

\lefthead{Cosmological implications of the MAXIMA-I CMB anisotropy measurement.}
\righthead{Stompor \etal}

\title{Cosmological implications of the MAXIMA-I high resolution Cosmic Microwave Background anisotropy measurement.}

\author {R.~Stompor$^{1,2,3}$, M.~Abroe$^{4}$,
P.~Ade$^{5}$, 
A.~Balbi$^{6}$, D.~Barbosa$^{7,8}$, J.~Bock$^{9,10}$,
J.~Borrill$^{11}$, A.~Boscaleri$^{12}$, P.~De Bernardis$^{13}$,
P.~G.~Ferreira$^{14}$, 
S.~Hanany$^{4}$, V.~Hristov$^{10}$, A.~H.~Jaffe$^{1,2,15}$, 
A.~T.~Lee$^{16,7,1}$, E.~Pascale$^{12}$, B.~Rabii$^{16,2}$, 
P.~L.~Richards$^{16,2}$, G.~F.~Smoot$^{16,7,2}$, 
C.~D.~Winant$^{16,2,1}$, J.~H.~P.~Wu$^{15}$}

\begin{abstract}
We discuss the cosmological implications of the new constraints
on the power spectrum of the  Cosmic Microwave Background Anisotropy 
derived from a new high resolution analysis of the \maximai\ measurement.
The power spectrum indicates excess power at $\ell \sim 860$ over
the average level of power at $411 \le\ell \le 785.$ This excess is
statistically significant on the $\sim 95$\% confidence
level. Its position coincides with that of
the third acoustic peak as predicted by generic inflationary
models, selected to fit the first acoustic peak as observed in the data.
The height of the excess power agrees with the predictions of a family of inflationary models with 
cosmological parameters that are fixed to fit the CMB data previously 
provided by \boom\ and \maximai\ experiments.
Our results, therefore, lend  support for inflationary models
and more generally for the dominance of adiabatic coherent perturbations in
the structure formation of the Universe.
At the same time, they seem to disfavor a large variety of the
non-standard (but inflation-based) models 
that have been proposed to improve the quality of fits to the CMB data
and consistency with other cosmological observables.

Within standard inflationary models, our results 
combined with the $COBE$-DMR data give best fit values and $95$\%
confidence limits for  
the baryon density, $\Omega_{\rm b}h^2\simeq 0.033{\pm0.013}$,
and the total density, $\Omega=0.9{+0.18\atop -0.16}$.
The primordial spectrum slope ($n_{\rm s}$) and the
optical depth to the last scattering surface ($\tau_{\rm c}$) are found to
be degenerate and to obey 
the relation $n_{\rm s} \simeq (0.99 \pm 0.14) + 0.46 \tau_{\rm c}$,
for $\tau_{\rm c} \le 0.5$ (all $95$\% c.l.).
\end{abstract}

\section{Introduction}
Observations of  Cosmic Microwave Background (CMB) temperature
anisotropy are reaching maturity. The high signal-to-noise
multi-frequency data gathered by the balloon-borne  \boom\ (de
Bernardis \etal 2000) and \maximai\ (Hanany \etal 2000) experiments
set stringent constraints on the shape of the power spectrum in the
broad range of angular scales ranging from $\sim 5^\circ$  down to
$\sim 10'$  scales (corresponding to a range in $\ell$-space from $\sim 50$ up to $\sim 600$ for \boom\ and from $\sim 35$ to
$\sim 800$ for \maximai. The measurements firmly established the
existence of a peak in the power spectrum at $\ell \sim 220$,
also suggested by the data from earlier\\
\vskip -0.1truecm
\vbox{{\scriptsize
\noindent$^{1}${Center for Particle Astrophysics, University of
  California, Berkeley, CA, USA}\\
$^{2}${Space Sciences Laboratory, University of California,
  Berkeley, CA, USA}
$^{3}${Copernicus Astronomical Center, Warszawa, Poland}\\
$^{4}${School of Physics and Astronomy, University of
  Minnesota/Twin Cities, Minneapolis, MN, USA}\\
$^{5}${Queen Mary and Westfield College, London, UK}\\
$^{6}${Dipartimento di Fisica, Universit\`a Tor Vergata,
  Roma, Italy}\\
$^{7}${Physics Division, Lawrence Berkeley National
  Laboratory, University of California, Berkeley, CA, USA}\\
$^{8}${CENTRA, Instituto Superior Tecnico, Lisboa,
  Portugal}\\
$^{9}${Jet Propulsion Laboratory, Pasadena, CA, USA}\\
$^{10}${California Institute of Technology, Pasadena, CA,
  USA}\\
$^{11}${National Energy Research Scientific Computing Center,
  Lawrence Berkeley National Laboratory, Berkeley, CA, USA}\\
$^{12}${IROE-CNR, Firenze, Italy}\\
$^{13}${Dipartimento di Fisica, Universit\`a La Sapienza,
  Roma, Italy}\\
$^{14}${Astrophysics, University of Oxford, UK}\\
$^{15}${Dept. of Astronomy, University of California,
  Berkeley, CA, USA}\\
$^{16}${Dept. of Physics, University of California,
  Berkeley, CA, USA}
}}\noindent
observations (\eg,\ Miller \etal 1999
and Mauskopf \etal 2000).
Because no secondary peaks were indisputably seen, there was
no unambiguous evidence for an inflation-like scenario where
structure formation is driven by passive adiabatic coherent fluctuations
(but see also results concurrent with ours -- Netterfield \etal 2001, and Halverson \etal
2001).

The level of power detected
beyond the first peak, on subdegree scales, was found to be somewhat
lower than that generally expected for inflation-based
``concordance'' models of pre-2000 (\eg, Ostriker \&\ Steinhardt 1995,
Krauss \&\ Turner 1995).
Although very good fits to the \maximai\ and \boom\ data could be found within the
inflationary family of models (\eg, Jaffe \etal 2001),
the most favored values of the baryon density were shown to be 
higher than that inferred from the standard arguments based on the
cosmological (Big Bang) nucleosynthesis (BBN), though the latter was found to be within the
$95$\% ($97$\%) confidence limits derived from the \maximai\ (\maximai+\boom) measurement.
Notwithstanding this fact a number of alternatives/extensions to the standard
cosmologies were suggested (\eg, Bouchet \etal 2000, Enqvist,
Kurki-Suonio, \&\ Valiviita, 2000,  Peebles, Seager, \&\ Hu, 2001).
Further high resolution data are required to test these models.

A companion \letter, Lee \etal (2001), presents a new analysis of
the \maximai\ data that extends the measured power spectrum from
the previously published range $35 < \ell < 785$ up to $\ell \le 1235$.  
In this \letter\ we discuss the cosmological significance of the new result.

This \letter\ is organized as follows.  In section~2, we look for the
signature of acoustic oscillations in our power spectrum,
the existence of which is predicted by inflation-motivated scenarios.
In section 3, we discuss the cosmological
implications of such a feature and derive constraints on some of the
cosmological parameters within the family of standard inflationary
cosmological models.  

\vbox{\vskip 0.cm \hskip -0.5cm\epsfxsize=9.cm\epsfbox{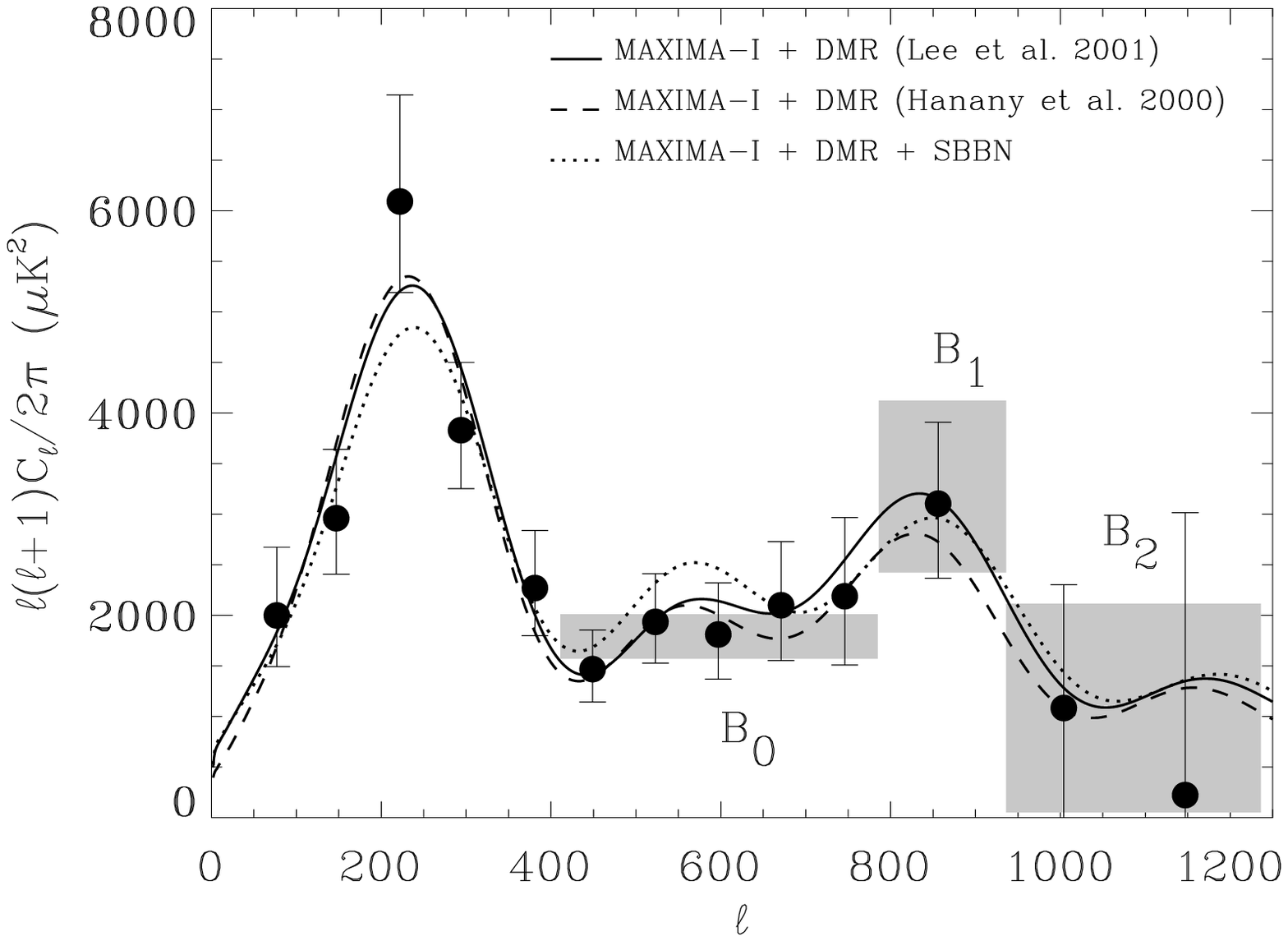}} 
\vskip 0.1cm
{ \small
  F{\scriptsize IG}.~1.--- Angular power spectra of the CMB anisotropy recovered from the high 
resolution map of the \maximai\ (Lee \etal 2001). The shaded
rectangles show the ranges of the bins used in our likelihood analysis
in Section~2, and the $68$\% confidence limits on the power level in these bins.
The dashed line shows the best-fit model to the previous low resolution
results of \maximai\ (Hanany \etal 2000), the solid line shows the
best-fit model for the new data (Lee \etal 2001), and the dotted line
shows the best-fit model, fulfilling the BBN constraint, found within the subfamily of all models considered here.
\label{fig:fig1}}
\vskip 0.25cm

Finally, in section 4  we present our conclusions
and comment briefly on the viability of non-standard cosmological
models in light of the new data.

\section{The excess power at $\ell \sim 860\pm75$}
Lee \etal (2001) set useful
constraints on the CMB anisotropy power spectrum up to $\ell \sim 1200,$
corresponding to angular scales down to $5'$. In terms of the simple $\chi^2$
statistics\footnote{Here and below, we define $\chi^2$
assuming a Gaussian distribution rather
than the approximately offset-lognormal form used in the explicit
likelihood calculations.},
this power, given as $\ell\l(\ell+1\r)C_\ell/2\pi$  in Figure~1, is consistent 
with a straight horizontal line for $\ell \simgt 410$ yielding a value $\chi^2\simeq 5.2$
for 7 degrees of freedom.
However, not only does such a model have no physical justification, but also
the data themselves qualitatively suggest the presence of high power 
at $\ell \sim 860$, in excess of the power level at $\ell\sim 410-785$,
and $\ell \sim 1000-1200$.
In fact a feature at $\ell \sim 850$, a third acoustic peak, is
anticipated within a family of generic
inflation-based models which are selected to reproduce a first peak at a position
$\ell \sim 220$ as seen in the data. A statistically
significant detection of such a feature would provide important additional support to
this family of models.
Statistical and systematic errors and
correlations between the spectral bands obscure the
statistical significance of the results.
We therefore employ a likelihood approach to 
assess the statistical importance of the feature at $\ell\sim860$ in the
\maximai\ power spectrum.

We work within a three-dimensional parameter space.
The initial parameters are the bin powers ${\cal C}_{0}$, ${\cal
C}_{1}$ and ${\cal C}_{2}$, for bins, $B_0$, $411 \le \ell \le 785$,
$B_1$, $786\le \ell \le 925$, and, $B_2$, $ 926\le \ell \le
1235$, as shown by the shaded regions in Figure~1.
Those are chosen to facilitate the search for the feature at $\ell \sim 850$.
Such a family of spectrum models includes a flat spectrum as a special
case, and in the Bayesian spirit our analysis seeks to
determine how well such a model represents our data.
\vskip 0.25cm
\vbox{\vskip 0.cm \hskip -0.5cm\epsfxsize=9.cm\epsfbox{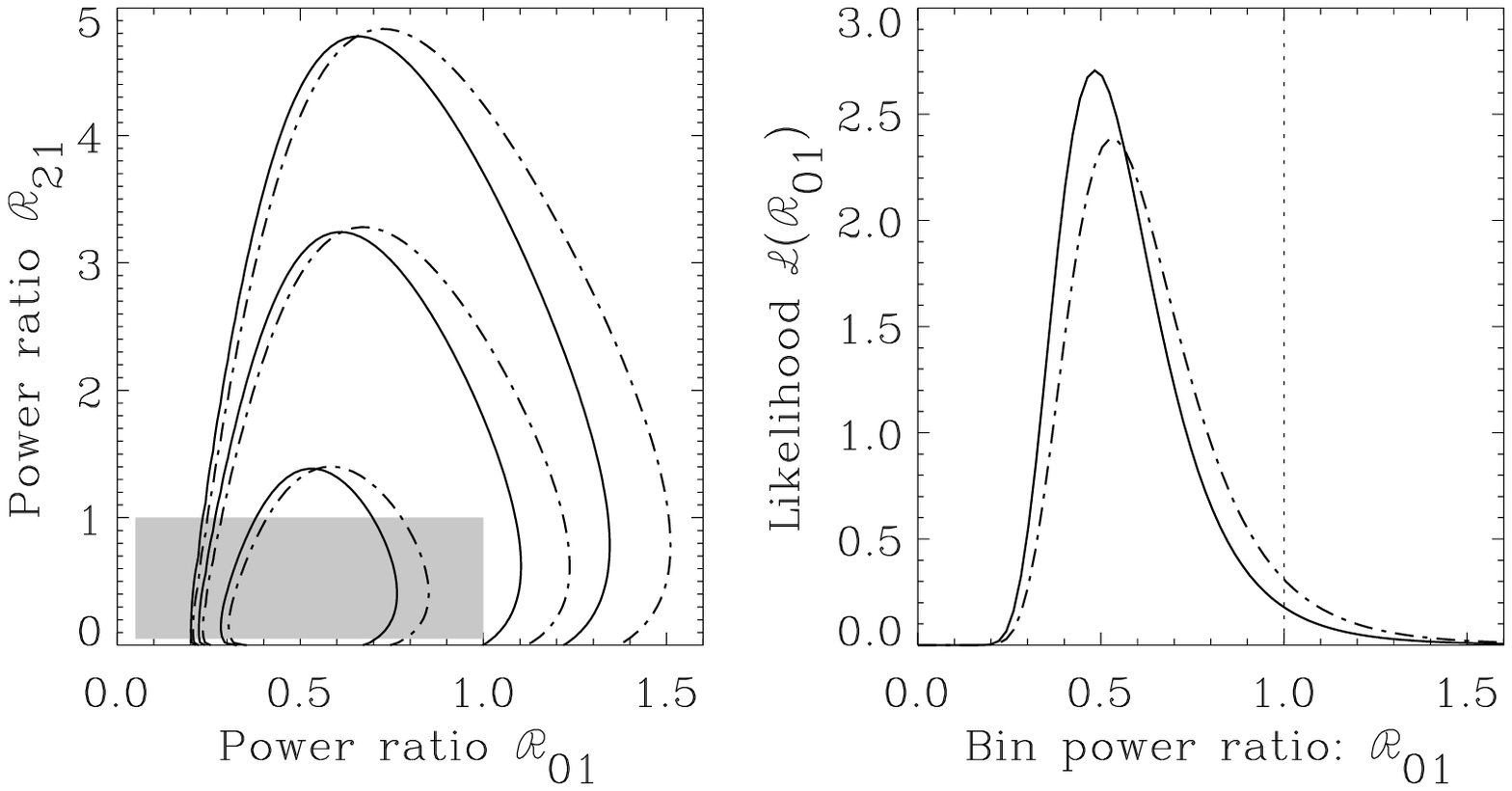}} 
\vskip -2.0cm
{ \small 
F{\scriptsize IG}.~2.--- {\it Left:} Two-dimensional likelihood as a function of the bin power ratios.
The contours correspond to $68,\ 95\ \&\ 99$\% confidence
levels computed through integration over the regions enclosed within
given contours. The shading
shows the area, which correspond to the existence of the ``excess
power'' in the bin centered at $\ell\sim 860$. 
{\it Right:} The one-dimensional likelihoods for bin power ratios ${\cal R}_{01}$.
Different lines are used for the two definitions of $B_0$. The solid lines are
for $411 \le \ell \le 785$ and the dashed lines for $486 \le \ell \le 785$.
}
\vskip 0.25cm

The likelihood for the three-dimensional parameter space is computed assuming the offset lognormal approximation 
of Bond, Jaffe \&\ Knox (2000) to the probability distribution of the
bin powers, denoted hereafter ${\cal L}\l({\cal C}_{0}, {\cal C}_{1}, {\cal C}_{2}\r)$.
The likelihood also depends on ``nuisance''
parameters describing the contribution of systematic effects such as 
the total calibration uncertainty and the beam and pointing
error. We model all of these as fully correlated between bins and Gaussian-distributed with a
dispersion depending on $\ell$ as given in Lee \etal (2001). 
We marginalize over these parameters by numerical integration.

The question we first ask is whether the power in the two side bins ($B_0$ and $B_2$) is lower
than the power in the middle bin centered at $\ell \sim 860,$ and at what
confidence level.
For that reason, we introduce two new parameters, ${\cal R}_{01}$ and ${\cal R}_{21}$,
which are given by
the ratio of power between bins ${\cal R}_{ij}\equiv {{\cal C}_{i}/{\cal C}_{j}}$.
As a third parameter, describing the overall normalization, we choose
the bin power amplitude defined above: ${\cal C}_{1}$. 
We adopt flat priors in the bin powers and marginalize over the
parameter, ${\cal C}_{1}$, leaving only a two dimensional problem,
\begin{equation}
{\cal L}\l({\cal R}_{01},{\cal R}_{21}\r)\propto \int {d{\cal
C}_{1}\; {\cal C}_{1}^2 }\; {\cal L}\l[{\cal C}_{0}\l({ \cal
R}_{01},{\cal C}_{1}\r),{\cal C}_{1},{\cal C}_{2}\l({\cal R}_{21},{\cal C}_{1}\r)\r],
\end{equation}
where ${\cal C}_1^2$ is a Jacobian. The results are shown in Figure~2. The left panel shows the
contours of the two dimensional likelihood ${\cal L}\l({\cal
R}_{01},{\cal R}_{21}\r)$ as a function of the ratios ${\cal R}_{01}$
and ${\cal R}_{21}$. The contours show $68,\ 95,\ 99$\% confidence
levels computed assuming flat priors for both parameters.
The shaded area indicates the region of parameter space favoring 
excess power in the middle bin relative to its
neighbors. Models without
excess power can not be rejected at a confidence level higher than ``$1\sigma$''
($\sim 68$\%). 

It is also apparent, however, that most of the models
preferred by the data, have power in the leftmost bin,
$B_0$, that is lower than that in the central bin, $B_1$.
This point is addressed in the right panel of 
Figure~2, showing the one-dimensional likelihoods computed through an
explicit marginalization over the other parameter.
We find that ${\cal R}_{01}=0.49{ +0.46\atop -0.21}$ and hence it is lower than $1$ at the confidence level
$\sim 95$\% (\cf, the value of ${\cal R}_{01}$ computed in the next section),
yielding a ``$2\sigma$'' detection
of the power rise at $\ell\sim 860$ over the intermediate-$\ell$-range
covered by the $B_0$ bin. 
This result remains unchanged if we allow for the presence of the
(uncorrelated) point-source-like component with the amplitude as expected
for the $150$ GHz band of \maximai\ even in the most
pessimistic cases (Lee \etal 2001).

It may appear that this high confidence level is largely due to the
leftmost point of the spectrum (at $\ell \sim 435$) included in our (clearly arbitrary)
definition of $B_0$. However, excluding that point
lowers the confidence level only to $\sim 93$\%.
Because of the large uncertainties of the Lee \etal results beyond
$\ell \simeq 900$ 
the statistical confidence that the power declines
beyond the bin $B_1$ is only $\sim 80$\%. However, the recent CBI 
result (Padin \etal 2001) constrains the power to $882{+663\atop
-428}\mu K^2$ at $\ell\simeq 1190 {+261 \atop - 234},$
providing extra and independent support to the presence of
such a decline. The high power level we see in the \maximai\ data at
$\ell \sim 860$, therefore can not extend far beyond the right edge of
the bin $B_1$ ($\simeq 925$).
The amplitude of this excess power is restricted by our data to 
${\cal C}_{1}= 3273{+1750\atop -1580} \mu K^2$
at $95$\% c.l. including both statistical and  systematic errors.

In the usual family of inflation-based models,
one also expects to find a second peak in the region of $410 \le \ell \le 785$.
While a single
flat band power provides an excellent fit to all the points within bin $B_0$
with $\chi^2\sim 1.5$ and $0.26$ for 4 (wider bin) and 3 (narrower) degrees of 
freedom\footnote{ Due to the strongly asymmetric shape of the $\chi^2$ distribution with so few
degrees of freedom; the quotoed numbers are very close to the maxima of those
distributions.}, the presence of a second peak in that region 
is still comfortably admissible by the data (Figure~1).
With our choice of the range of $B_0$, this bin is expected to be only
a factor $\simlt 2$ wider than a typical feature of the
power spectrum ($\Delta \ell \sim 150$) of the standard inflationary
model. If such a peak structure is present in that range, 
$R_{01} < 1$ means that the bin power at $\ell\sim 860$ is higher than the level of power in 
some subsection of the $410\le\ell\le786$ range. Then our analysis
above demonstrates that this happens at a
confidence level of {\it at least} $95$\% and that value is our
confidence level of detecting a 
feature in the power spectrum at $\ell\sim 860$.

\section{Cosmological parameters}

\subsection{Parameter space}

Figure~1 demonstrates that an inflationary model with 
parameters as determined by fitting to the \maximai\ power spectrum published by Hanany \etal (2000) provides a very
good fit over the entire current range of our new 
data together with the $COBE$-DMR result. The total $\chi^2$ is $\simeq 32$ for 41 data points (\maximai\ plus DMR) and $\chi^2\simeq 8$ for the
13 points of \maximai\ only.
Thus, our new data are consistent with the predictions made on the
basis of the relatively narrow class of inflation-motivated models
assumed in the previous papers.
Such a class of models can be seen as a particular subset of
the family of models considered in the previous Section.

 Here we consider a seven-dimensional space of parameters. The parameters include the
 amplitude of fluctuations at $\ell=10$, $C_{10}$, the 
 physical baryon density, $\Omega_{\rm b}h^2$,
the physical density of cold dark matter, $\Omega_{\rm cdm}h^2$, 
 and the cosmological constant, $\Omega_\Lambda$, 
 the total energy density of the universe, 
 $\Omega\equiv \Omega_{\rm b}+\Omega_{\rm cdm}+\Omega_\Lambda$,   
 the spectral index of primordial scalar fluctuations, $n_{\rm s}$, and the
 optical depth of reionization, $\tau_{\rm c}$. We use the following ranges
 and sampling: $C_{10}$ is continuous; $\Omega$ = 0.3, 0.5, 0.6, 0.7, 0.75, ..., 1.2, 1.3, 1.5;
$\Omega_{\rm b}h^2$ = 0.00325, 0.00625, 0.01, 0.015, 0.02, 0.0225, ..., 0.04, 0.045, 0.05, 0.075, 0.1;
$\Omega_{\rm cdm}h^2$ = 0.03, 0.06, 0.12, 0.17, 0.22, 0.27, 0.33, 0.40, 0.55, 0.8;
$\Omega_\Lambda$ = 0.0, 0.1, 0.2, ..., 1.0;
$n_{\rm s}$ =
0.6, 0.7, 0.75, 0.8, 0.85, 0.875, ..., 1.2, 1.25, ..., 1.5;
$\tau_{\rm c}$ = 0, 0.025, 0.05, 0.075, 0.1, 0.15, 0.2, 0.3, 0.5.
The justification of the presented choice of the parameter space can
be found in Balbi \etal 2000.
Models were computed using a version of {\sc cmbfast} by Tegmark, M., Zaldarriaga, M., \&\ Hamilton, A., (2001),
originally by Seljak \&\ Zaldarriaga (1996).

\subsection{Results}
We compute the likelihood on the grid for models using an offset
lognormal approximation (Bond, Jaffe, \&\ Knox 2000), 
including both statistical and systematic errors in a manner analogous to Sect.~2.
We neglect a subdominant pointing uncertainty (Lee \etal 2001).
The likelihood for a subset of parameters is evaluated by an explicit marginalization
over all the remaining parameters, using a top-hat prior in parameters
used to define our database. In addition, we impose top-hat priors for
the value of the Hubble constant $0.4\le h \le 0.9$, the matter density,
$\Omega_{\rm m} \equiv\Omega_{\rm b}+\Omega_{\rm cdm}> 0.1$ and the
age of the Universe $>10$~Gyr.
We marginalize over the calibration and beam uncertainty in order
to account for remaining systematic uncertainties in our results.
To provide an extra large-angular-scale constraint we combine our
data with the results from the $COBE$-DMR satellite as provided by G\'orski \etal (1996).

Within the chosen family of inflationary models we put stringent 95\% confidence
level constraints on the total density, $\Omega=0.9 {+0.18 \atop -0.16}$,
baryon density, $\Omega_{\rm b} h^2 =0.033\pm0.013$
and power spectrum normalization $C_{10} = 690 {+200\atop -125} \mu K^2$.
Our $95$\% confidence limit on the cold dark matter density is 
$\Omega_{\rm cdm}h^2=0.17{+0.16\atop -0.07}$.
However, this result is mostly determined by the priors defining the 
database parameter range, as discussed in Jaffe \etal (2001).
We find a strong degeneracy between the optical depth 
to the last scattering surface, $\tau_{\rm c},$ and the primordial power
spectrum index, $n_{\rm s}$. In this case, the degeneracy 
restricts the parameters to a subspace allowing us to derive a combined constraint: $n_{\rm s}
\simeq (0.99 \pm 0.14) + 0.46\tau_{\rm c}$
($95$\% c.l.), for $\tau_{\rm c} \le 0.5$. We note that we recover the
$95$\% upper limit ($\simlt 0.4$) on $\tau_{\rm c}$, derived by
Griffiths, Barbosa, \&\ Liddle (1999), 
if we constrain the spectral index to be $\le 1.2$ as was assumed by
those authors. 
Independent of the value of the optical depth, we can put a very firm $\simgt 99$\% lower
limit on the spectral index, $n_{\rm s}\ge 0.8$.
Assuming no reionization ($\tau_{\rm c}=0$), we can get both lower and upper
limits on the spectral index reading $n_{\rm s}=0.99\pm0.14$. Alternatively, fixing
$n_{\rm s}$ at unity gives us a $95$\% c.l. upper limit $\tau_{\rm c}\simlt 0.26$.

The $\chi^2$ of the best fit model is $30$ for all 41 points
used in the fittings, and $4$ for the 13 points of \maximai\ only.
The best fit model parameters $(\Omega_{\rm b},\Omega_{\rm cdm},\Omega_\Lambda,\tau_{\rm c},n_{\rm s},h) = 
\l(0.07,0.68,0.1,0.0,1.025,0.63\r)$ are
characterized by high matter and low vacuum energy content (see also Balbi \etal 2000).
However, due to strong degeneracy between $\Omega_{\rm m}$ and $\Omega_\Lambda$ 
(\eg, Zaldarriaga, Spergel \&\ Seljak 1998) we can easily find models comfortably fulfilling both \maximai+DMR and supernovae 
constraints (see Figure~3). 
These data sets constraint jointly the
matter and vacuum energy densities to be: $(\Omega_{m},
\Omega_{\Lambda})=(0.32{+0.14\atop -0.11}, 0.65{+0.15\atop -0.16})$ (95\% c.l.).

The constraint on $\Omega_{\rm b} h^2$ mentioned above is compatible with the best determination to date of the baryon density based on
measurements of primordial deuterium and calculations of standard BBN (Burles, Nollett, \&\ Turner, 2001,
Tytler \etal 2000), $\Omega_{\rm b} h^2=0.020 \pm 0.002$. 
The consistency of the data with BBN becomes even more apparent by
constraining our parameter estimation on the BBN value of
$\Omega_{\rm b}h^2,$ which we approximate by fixing $\Omega_{\rm b} h^2$ at the grid
value nearest to the BBN prediction (\ie, $\Omega_{\rm b} h^2=0.02$).
The best fit model then has parameters $(\Omega_{\rm b},\Omega_{\rm cdm},\Omega_\Lambda,\tau_{\rm c},n_{\rm s},h) = \l(0.07,0.78,0.0,0.0,1.0,0.53\r)$ 
with $\chi^2 \simeq 7$ 
for the 13 \maximai\ points only. This very good fit emphasizes the
compatibility of our data with other cosmological measurements. 
The spectrum of this model is indicated in Figure~1 with a dotted line,
and it shows that even in that case,
the best fit model has a higher third than second peak, if $\ell
(\ell+1)C_\ell$ is plotted versus $\ell$.
\vskip -3.95cm
\vbox{\vskip 0.cm \hskip -1.5cm\epsfxsize=8.0cm\epsfbox{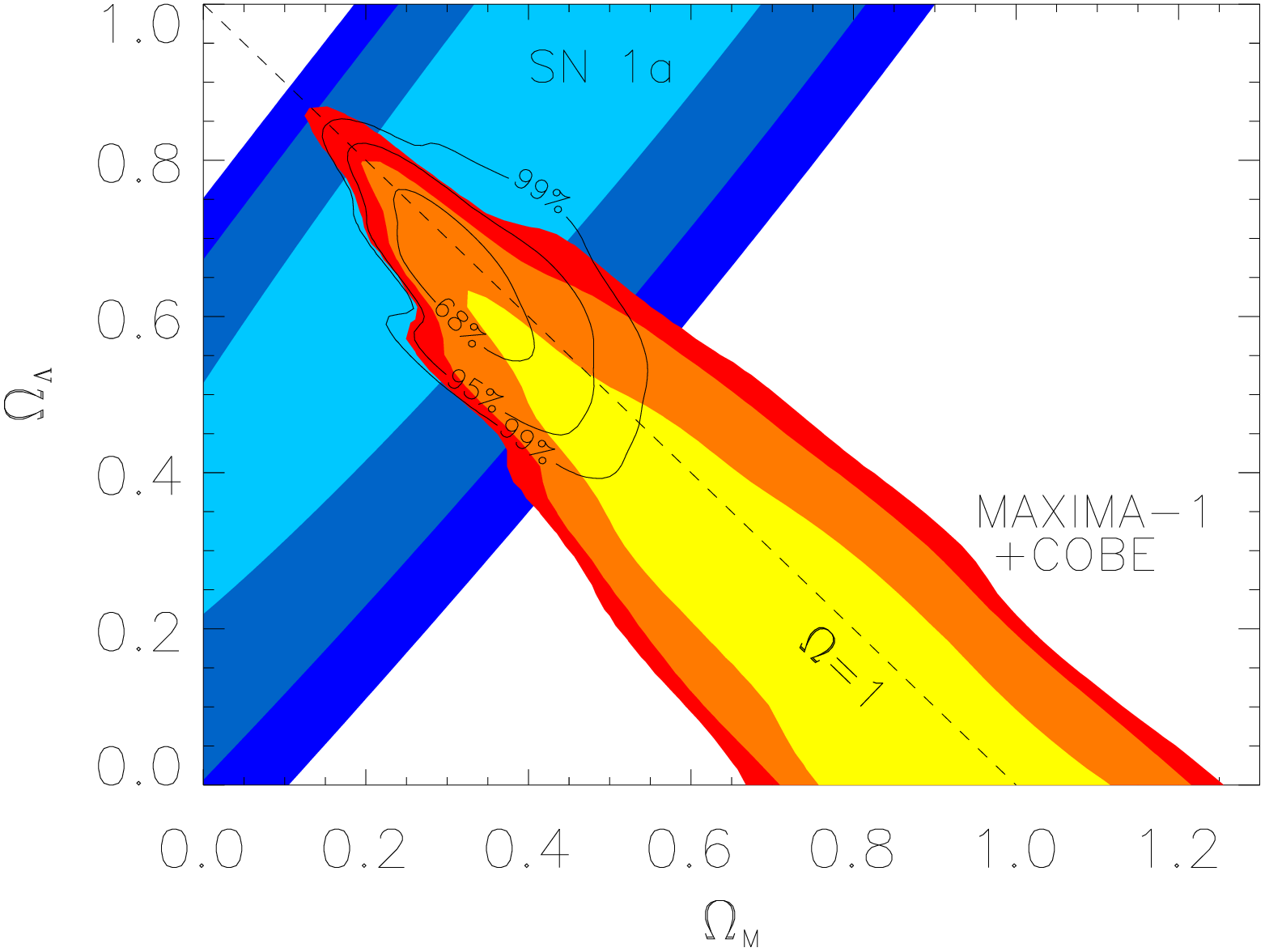}} 
\vskip 0.5cm
{ \small
  F{\scriptsize IG}.~3.--- 
  Constraints in the $\Omega_{\rm m}$--$\Omega_\Lambda$ plane from the
  combined \maximai\ and $COBE$-DMR datasets. The shown contours
  correspond to $68,\ 95,\ \&\ 99$\% likelihood-ratio confidence levels.
  The  bounds obtained from high redshift supernovae data (Perlmutter \etal
  1999; Riess \etal 1998) are also overlaid as well as 
  the confidence levels of the joint likelihood. 
\label{fig:fig3}}
\vskip 0.5cm
We also compute constraints on the ratio of bin powers, ${\cal R}_{01}$,
as imposed by the \maximai\ data within the discussed family of models.
We find that the most likely value of that ratio is $0.68 {+0.27\atop -0.13}$ at
the $95$\% confidence level, consistent with (though on average higher than) our 
result in Sect.~2 (\cf, Figure~2).

\section{Implications for cosmological models}

Inflation-based models provide us with an abundance of excellent fits to the \maximai\
extended power spectrum. These include the best fit models found by the analysis of the first data sets
of \boom\ and \maximai\ (Jaffe \etal 2001, Balbi \etal 2000). Constraints on cosmological parameters
using DMR and \maximai\ confirm the near flatness of the Universe and,
when combined with recent supernova data (Riess \etal 1998,
Perlmutter \etal 1998), support the need for the non-zero vacuum energy density. The most likely baryon density indicated 
by our data is found to be somewhat higher than the preferred BBN
value, yet the latter is 
within the $95$\% confidence range of our determination. Moreover,
there are excellent fits to the data with the baryon fraction at the BBN value.

Our results constrain the power at $\ell \sim 860$ to be 
$\simgt 1700\mu K^2$ on the $95$\% confidence level.
They also
indicate an increase in the power spectrum between $\ell \sim 410-785$ and $\ell \sim 786-925$, yielding
the ratio of the corresponding bin powers equal to ${\cal R}_{01}=0.49
{+0.46\atop -0.21}$ in a general case,  or ${\cal R}_{01}=0.68 {+0.27\atop -0.13}$
within the considered family of inflationary models.
We have shown that such constraints can be easily fulfilled by these
inflationary models. However, they impose strong requirements on some
non-standard models which were found to better accommodate the first results of the \boom\ and \maximai\ experiments.
Those models, by design, have an amplitude of power in the
intermediate range of $\ell$ below the typical predictions of the
standard inflationary models.
However, they also tend to have lower power at the high $\ell$ end 
and therefore, to match our new constraint at $\ell \sim 860$,
they would require even higher baryon abundance than the family of the
models considered in Sect.~3.
These disfavored models include: the mixture of inflation and topological defect models (Bouchet \etal 2000, Contaldi 2000), 
hybrid adiabatic plus isocurvature models (Stompor, Banday, \&\ G\'orski, 1996, Enqvist, \etal 2000), 
broken power-law primordial spectra models 
(Griffiths, Silk, \&\ Zaroubi, 2001, Barriga \etal 2001), or models with a delayed recombination (Peebles \etal 2001).
Further investigation is required to fully elucidate status of these models.
\acknowledgments
This work was supported by NASA grants no.\ NAG5-3941 (RS, SH),
NAG5-6552 (JHPW, AHJ), S00-GSRP-032 and S00-GSRP-031 (BR, CDW);
NSF KDI Grant no.\ 9872979 (JHPW, AHJ); and the Royal Society (PGF).
Computing resources
were provided by the Minnesota Supercomputing Institute and the
National Energy Research Computing Center at Lawrence Berkeley
National Laboratory.
MAXIMA is supported by NASA Grant
NAG5-4454 and by the NSF through the Center for Particle
Astrophysics at UC Berkeley, NSF cooperative agreement AST-9120005.
The likelihood from SN data was kindly provided by S. Jha and the High-Z SN Search Team.

\end{document}